\documentclass[12pt]{report}
\usepackage{psfig}

\begin{document}
\pagestyle{myheadings}
\markright{Manuscript \#2004-01-14754}

\noindent{\bf \large The geometry of the double-pulsar system J0737-3039 from systematic intensity variations.}

\noindent Fredrick A. Jenet$^1$ \& Scott M. Ransom$^{2,3}$

%\vspace{12pt}

\noindent $^1${\em California Institute of Technology, Jet Propulsion
Laboratory, 4800 Oak Grove Drive, Pasadena, CA 91109;
merlyn@alum.mit.edu}

%\vspace{12pt}

\noindent $^2$ {\em McGill University Physics Dept., Montreal, QC H3A
2T8, Canada; \\ransom@physics.mcgill.ca}

%\vspace{12pt}

\noindent $^3$ {\em Center for Space Research, Massachusetts Institute
of Technology, Cambridge, MA 02139}

\noindent {\bf The recent discovery of J0737-3039A$^1$ \& B$^2$-two pulsars
in a highly relativistic orbit around one another - offers an
unprecedented opportunity to study the elusive physics of pulsar radio
emission. The system contains a rapidly rotating pulsar with a spin
period of 22.7 ms and a slow companion with a spin period of 2.77 s,
hereafter referred to as `A´ and `B´, respectively. A unique property
of the system is that the pulsed radio flux from B increases
systematically by almost two orders-of-magnitude during two short
portions of each orbit$^2$.  Here, we describe a geometrical model of the
system that simultaneously explains the intensity variations of B and
makes definitive and testable predictions for the future evolution of
the emission properties of both stars.  Our model assumes that B's
pulsed radio flux increases when illuminated by emission from A. This
model provides constraints on the spin axis orientation and emission
geometry of A and predicts that its pulse profile will evolve
considerably over the next several years due to geodetic precession
until it disappears entirely in 15-20 years.}

The double-pulsar system is extremely compact (orbital period,
$P_{orb} \approx 2.45$h), mildly eccentric ($e \approx 0.088$), and
highly inclined ( $i > 87^o$).  The orbital light curve shows two
regions that last for about 12 min ($30^o$ of orbital phase) and are
separated by approximately 74$^o$ where the pulsed radio flux of B
increases dramatically$^2$. The average pulse profile of A is composed
of two peaks of width $20^o-30^o$ separated by about $150^o$
in pulse phase$^1$.

We report on a model that explains the pulsed intensity variations
seen in the orbital light curve of B using two simple
assumptions. First, we assume that the emission geometry of A is given
by the traditional hollow cone model of Lyne \& Manchester$^3$ (see
figure 1), and second, that B appears bright only when the emission
beam of A illuminates it (i.e. when the cone of emission from A
intersects B's position in the orbital plane).  The key feature of the
second assumption, which we call the "stimulated emission hypothesis",
is that the particles and/or electromagnetic radiation from A which
stimulate the radio emission from B are directed along the same
emission cone as A's observed radio radiation. These two assumptions
enable one to derive a set of non-linear equations (see the caption to
figure 1) that determines how the conal emission from A intersects
with our line-of-sight, how it projects onto the orbital plane, and
when it illuminates B. These five equations contain a total of 11
variables, six of which are obtained from observations, and which
therefore allow us to solve for the five unknowns. The six measured
parameters are the orbital inclination angle, $i$, the angle between
the outer edges of the double peaked pulse profile, $P_o$, the angle
between the inner edges of the peaks in the pulse profile, $P_i$, the
angle between the outer edges of the double peaked orbital light
curve, $O_o$, the angle between the inner edges of the orbital light
curve, Oi, and the midpoint of the orbital light curve, $O_m$. The
angles obtained by solving the five constraint equations are $\lambda$
and $\phi$ which are the polar angles that determine the orientation
of the spin axis with respect to the orbital angular momentum (see
figure 1), and $\rho$, $\alpha$, and $\delta$, which determine the
emission geometry (see figure 2).  

Table 1 lists the adopted values of the measured parameters $i$,
$P_i$, $P_o$, $O_i$, $O_o$, and $O_m$. These values were obtained from
427, 820, 1400, and 2200 MHz data taken with the Green Bank
Telescope$^4$. These observations, which cover a factor of 5 in radio
frequency, indicate that the pulse profile of A and the orbital light
curve of B are both effectively independent of observing frequency for
our purposes. Given the measured parameters, we solved the constraint
equations and found two solution sets for the five unknown angles,
which are also listed in Table 1. These two solutions exist due to the
uncertainties in the measured parameters. A video showing the relative
motions of the pulsars, the intersection of A's conal emission with
the orbital plane for solution 1, and the regions of enhanced emission
from B is available on-line$^5$.  

In our model, the orientation of A's spin axis plays a major role in
determining the shape of both the pulse profile of A and the orbital
light curve of B. Geodetic precession$^{6,7}$, caused by the curvature of
space-time around the pulsars, will cause the spin axis to precess
about the orbital angular momentum vector, {\bf J}, at a rate of
$4.77^o$ per year. Hence, the angle $\phi$ will increase at this rate
while the four other inferred angles will remain
constant. Consequently, we can predict the future evolution of both
the orbital light curve and the pulse profile of A. Given the inferred
values of $\lambda$, $\phi$, $\rho$, $\alpha$ and $\delta$ together
with a value of $\phi$ at a specified time, the five constraint
equations determine the pulse profile parameters, $P_i$, and $P_o$,
and the orbital light curve parameters $O_o$, $O_i$, $O_m$. Both $O_i$
and $O_o$ will remain constant while $O_m$, which is equal to $\phi$,
will increase at a rate of $4.77^o$ per year. This effect will be
easily measurable over the next 1-2 years. The geodetic precession
will also cause highly significant pulse profile changes over
timescales of a few years (see figure 3). Note that similar, although
smaller amplitude, precession-induced profile changes have already
been observed in the binary system PSR B1913+16$^{8,9}$. For solution 1,
$P_o$ and $P_i$ will increase by $42^o \pm 16^o$ and $31^o \pm 8^o$ in
1 year , respectively. For solution 2, $P_o$ and $P_i$ will increase
by $96^o \pm 13^o$ and $34^o \pm 6^o$ in one year, respectively. For
both solutions, $P_o$, will approach its maximum extent, 3600, in less
than 2 years. Pulsar A is expected to disappear in about 14 years
according to solution 1. For solution 2, the pulsar will disappear in
approximately 4.5 years and then reappear in 10 years as a
single-peaked pulsar (temporarily) where it will remain "on" for 6-7
years. For both solutions, we estimate that the pulsar came into view
approximately 4-5 years in the past. This can explain why the Parkes
70-cm survey for pulsars, completed in 1997, did not detect this
system in spite of sufficient sensitivity and appropriate sky
coverage$^{10}$.

Recently, Demorest et al.$^{11}$ used the polarization properties of A
together with the standard rotating vector model of Radhakrishnan \&
Cooke$^{12}$ in an attempt to measure its emission geometry. Due to
limitations in both the model and the data, they were only able to
measure the angle $\alpha$ which they found to be $4^o \pm 3^o$. This
result is consistent with our solution 1. Since they only measured one
of the five angles needed to completely specify the geometry, they
were unable to predict the future evolution of the emission properties
of this system.  

We note that the current form of the model does not consider the
precession of B. However, it is likely that wind-torques from A, which
dominates the system energetically, have caused the spin axis of B to
align with the direction of the orbital angular momentum$^{13,14}$.  In
this case, geodetic precession will have no effect on the emission
from B. It is also possible that the direction of B's emission beam
may lie in the orbital plane as a result of the stimulated emission
process regardless of the magnetic field alignment. In this scenario,
geodetic precession would also have little effect on the direction of
B's emission.  

Given that the model presented here accurately describes the current
data, it becomes important to understand the physics behind the
stimulation process. Future work will explore various possible
mechanisms. One idea involves "jump-starting" the pulsar emission
processes in B by initiating electron-positron pair cascades in its
magnetosphere that emit coherent radio emission.  The initiating
particles could be the positrons and electrons emitted in the wind of
A, or, more likely, gamma rays that are expected to be travelling in
nearly the same direction as the radio photons.  Alternatively,
pressure from a conal A-wind could distort B's magnetosphere$^{14}$ and
push its beam more directly into our line of sight.

\pagebreak

\begin{enumerate}
\item Burgay, M., D'Amico, N.,  Possenti, A.,  Manchester, R. N., Lyne, A. G.,  Joshi, B. C., McLaughlin, M. A., Kramer, M.,  Sarkissian, J. M.,  Camilo, F.,  Kalogera, V., Kim, C., \& Lorimer, D. R. An increased estimate of the merger rate of double neutron stars from observations of a highly relativistic system. Nature, {\bf 426}, 521-533 (2004)

\item Lyne, A. G., Burgay, M., Kramer, M., Possenti, A., Manchester, R. N.,  Camilo, F., McLaughlin, M. A., Lorimer, D. R., D'Amico, N., Joshi, B. C., Reynolds, J., Freire, P. C. C.,  A Double-Pulsar System - A Rare Laboratory for Relativistic Gravity and Plasma Physics. Science, {\bf 303}, 1153-1157 (2004)

\item Lyne, A. G., Manchester, R. N., The shape of pulsar radio beams.  Mon. Not. R. Astron. Soc., {\bf 234}, 477-508 (1988)

\item Ransom, S., Kaspi, V., Demorest, P., Ramachandran, R., Backer, D., Pfahl, E., Ghigo, F., Arons, J., Kaplan, D. Green Bank Telescope Measurement of the Systemic Velocity of the Double Pulsar Binary J0737-3039 and Implications for its Formation, Astrophys. J., submitted (2004)

\item $http://www.physics.mcgill.ca/~ransom/0737\_Bflux\_model.mpg $

\item Barker, B. M. \& O'Connell, R. F., Gravitational two-body problem with arbitrary masses, spins, and quadrupole moments, Phys. Rev. D., {\bf 12}, 329-335 (1975)

\item Barker, B. M. \& O'Connell, R. F., Relativistic effects in the binary pulsar PSR 1913+16, Astrophys. J., {\bf 199}, L25-L26 (1975) 

\item Kramer, M., Determination of the Geometry of the PSR B1913+16 System by Geodetic Precession, Astrophys. J., {\bf 509}, 856-860 (1998) 

\item Weisberg, J. M. \& Taylor, J. H., General Relativistic Geodetic Spin Precession in Binary Pulsar B1913+16: Mapping the Emission Beam in Two Dimensions, Astrophys. J., {\bf 576}, 942-949 (2002) 

\item Lyne, A. G., Manchester, R. N., Lorimer, D. R., Bailes, M., D'Amico, N., Tauris, T. M., Johnston, S., Bell, J. F., Nicastro, L., The Parkes Southern Pulsar Survey - II. Final results and population analysis., Mon. Not. R. Astron. Soc., {\bf 295}, 743-755 (1998)

\item Demorest, P., Ramachandran, R., Backer, D. C., Ransom, S. M., Kaspi, V., Arons, J., Spitkovsky, A.,  Orientations of Spin and Magnetic Dipole Axes of Pulsars in the J0737-3039 Binary Based on Polarimetry Observations at the Green Bank Telescope. Astrophysics J., submitted,  astro-ph/0402025 (2004)

\item Radhakrishnan, V., Cooke, D. J., Magnetic Poles and the Polarization Structure of Pulsar Radiation.  Astrophys. J., {\bf 3},  L225-L228 (1969)

\item Kaspi, V. M., Ransom, S. M., Backer, D. C., Ramachandran, R., Demorest, P.,  Arons, J., Spitkovskty, A.,  Green Bank Telescope Observations of the Eclipse of Pulsar B in the Double Pulsar Binary PSR J0737-3039.  Astrophys. J., submitted, astro-ph/0401614 (2004)

\item Arons, J., Spitkovskty, A., Backer, D., Kaspi, V. Probing Relativistic Winds and Magnetospheres: Torques and Eclipses of PSR J0737-3039 A \& B. Astrophys. J., submitted (2004)

\item Press, W. H., Flannery, B. P., Teukolsky, S. A. \& Vetterling, W. T. {\em Numerical Recipes in C. The Art of Scientific Computing} (Cambridge Univ. Press, 1988).
\end{enumerate}

\noindent {\small {\bf Correspondence} and requests for materials should be addressed to F. J. (merlyn@alum.mit.edu)}

\noindent {\small {\bf Competing interests statement} The authors declare that they have no competing financial interests.

\noindent {\small {\bf Acknowledgements} Part of this research was
performed at the California Institute of Technology's Jet Propulsion
Laboratory, under contract with NASA and funded through the internal
Research and Technology Development Program. The authors wish to thank
Vicky Kaspi, John Armstrong, Jonathan Arons, Donald Backer, Paulo
Freire, and Dick Manchester for useful discussions and
comments. F. J. extends special thanks to Thomas A. Prince and
E. B. Dussan V.}

\pagebreak

\noindent {\bf {Table 1: Measured and Inferred Properties of J0737-3039}}

\begin{tabbing}
{\bf Measured Parameters}\\
*******************************\=******************\=*******************\= \kill
Orbital Inclination        \> $i$            \> $88.5^o \pm 1.5^o$ or $91.5^o \pm 1.5^o$\\
Profile outer width	\> $P_o$	\>	$251^o \pm 10^o$\\
Profile inner width	\>	$P_i$	\>	$109^o \pm 10^o$ \\

Orbital outer width	\>	$O_o$	\>	$110^o \pm 10^o$ \\

Orbital inner width	\>	$O_i$	\>	$22^o \pm 10^o $ \\

Orbital Midpoint	\>	$O_m$	\>	$246^o \pm 10^o$ \\
\\
{\bf Inferred Parameters for $i=88.5 \pm 1.5$} \>	\> Solution 1  \> Solution 2\\
Spin Axis		\>	$\lambda$	\>	$167^o \pm 10^o$  \>	$90^o \pm 10^o$\\
			\>	$\phi$	\>	$246^o \pm 5^o$   \>	$239^o \pm 2^o$\\
Magnetic field angle	\>	$\alpha$	\>	$1.6^o \pm 1.3^o$ \>	$14^o \pm 2^o$\\
Emission ring opening angle\>	$\rho$	\>	$78^o \pm 8^o$    \>	$42^o \pm 4^o$\\
Emission ring width	\>	$2\delta$	\>	$1.9^o \pm 1.4^o$ \>	$15^o \pm 2^o$\\
Total visible time	\>              \>	$19 \pm 2$ yrs  \>	$19 \pm 2$ yrs\\
Time remaining		\>\>			$14 \pm 2$ yrs  \>	$4.5 \pm 0.1$ yrs\\
{\bf Predicted changes in 1 year} \>\>              Solution 1  \>       Solution 2\\
Change in $P_o$           		\>\>		    $42^o \pm 16^o$	\> $96^o \pm 13^o$\\
Change in $P_i$			\>\>		    $31^o  \pm 8^o$	\> $34^o  \pm 6^o$\\
Change in $O_o$			\>\>		     $ 0^o$	\>	  $0^o$\\
Change in $O_i$			\>\>		      $0^o$        \>         $0^o$\\
Change in $O_m$            	\>\>		$4.77^o \pm .01^o$  \>   $4.77^o \pm .01^o$

\end{tabbing}

Note - The measured and inferred geometric parameters for PSR
J0737-3039. Timing$^2$ and scintillation$^4$ observations have determined $i$
to be either $88.5^o \pm 1.5^o$ or $91.5^o \pm 1.5^o$. Two values are allowed
since current data are unable to distinguish between $i$ and $180^o-i$,
although future timing observations should resolve this degeneracy.
If a solution to the constraint equations exists for a given
inclination, $i$, then a solution will also exist for an inclination of
$180^o-i$. The only difference between the two solutions will be that $\lambda$
will become $180^o-\lambda$. In the text, the results are presented only for
the case $i < 90^o$ with the understanding that corresponding solutions
also exist if $i > 90^o$. Note that the evolution of a given solution
does not change when $\lambda$ goes to $180^o - \lambda$. We estimated the errors in
the inferred parameters by allowing the measured parameters ($i$, $P_o$,
$P_i$, $O_o$, $O_i$, $O_m$) to vary by the given uncertainties and then
calculating the variations in the resulting solutions. The solutions
were found by solving the five non-linear constraint equations using
Broyden's method$^{15}$. An initial guess for each of the five angles was
chosen at random and then refined using this algorithm. This was
performed 10000 times in order to determine all possible solutions.

\pagebreak

\noindent {\bf Figure 1} - The hollow cone model of radio pulsar emission. {\bf
$\Omega$} is the spin axis of the pulsar, which is separated by the
magnetic dipole axis {\bf $\mu$} by the angle $\alpha$. The conal
emission centered on {\bf $\mu$} has an opening half-angle $\rho$ and is of
angular thickness $2\delta$. The cut of an observer's line-of-sight through
the cone as it rotates can produce either a single or double-peaked
pulse profile. This emission geometry together with the stimulated
emission hypothesis yield the following five constraint equations:

\begin{eqnarray}
\sin(\lambda)\cos\left(\frac{O_i}{2}\right) &=& \left\{ \begin{array}{ll}
                                   \cos(\rho + \delta - \alpha) & ;\alpha  \ge \rho + \delta \\
                                   \sin(\lambda) & ;\rho - \delta < \alpha < \rho + \delta\\
                                   \cos(\rho - \delta - \alpha) & ;\alpha \le \rho - \delta
\end{array}  \right.\label{o1}\\
\sin(\lambda)\cos\left(\frac{O_o}{2}\right) &=& \cos(\rho + \delta + \alpha) \label{o2}\\
\phi &=& O_m\\
\cos(\rho - \delta) &=& A \cos(\alpha) + \sqrt{1 - A^2} \sin(\alpha)\cos(P_i/2)\\
\cos(\rho + \delta) &=& A \cos(\alpha) + \sqrt{1 - A^2} \sin(\alpha)\cos(P_o/2)
\end{eqnarray}

Here, $A=\cos(\lambda)\cos(i) -
\sin(i)\sin(\lambda)\sin(\phi)$. $P_i$, $P_o$, $O_i$, $O_o$, $O_m$,
and $i$ are measured parameters defined in the text.  $\lambda$ and $\phi$ are
defined in figure 2.
\\
\noindent {\bf Figure 2} - The geometry of key orbital- and
spin-related angles in the J0737-3039 system.  Orthogonal axes
representing the three spatial dimensions are denoted by {\bf x} (the
direction of the Line of Nodes in the orbital plane), {\bf y} (the
projected direction of the Line of Sight of an Earth-bound observer
onto the orbital plane), and {\bf z} (the direction anti-aligned with
the angular momentum vector of the orbit, {\bf J}).  The pulsars orbit
in the {\bf x-y} plane in a clockwise direction. The orientation of
the spin axis of J0373-3039A is {\bf $\Omega$}, which is separated by
the orbital angular momentum, {\bf J}, by the angle $\lambda$, and the
magnetic dipole axis, {\bf $\mu$}, by the angle $\alpha$. The
projection of {\bf $\Omega$} onto the orbital plane defines the angle
$\phi$. The line-of-sight of the observer is inclined by the
inclination angle $i$ from {\bf J} in the {\bf y-z} plane.  
\\
\noindent {\bf Figure 3} - A greyscale plot of the A pulsar pulse
profile evolution as a function of time for both sets of
solutions. The profile is given by a constant time slice. T=0 refers
to January 1, 2004 (MJD= 53005.0). Since this model only determines
the locations of the emission regions as opposed to the exact shape of
the profile, the simulated pulse profiles are represented as square
waves. We predict that the 'A' pulsar will disappear from view in
about 14 years for solution 1 or about 4.5 years for solution 2. The
lack of pulsed emission from either solution can explain the lack of
detection of the pulsars during the Parkes 70-cm pulsar survey in the
mid-1990s$^6$. Note that we have only considered one emission cone
emanating from one side of the A pulsar. Given the symmetries of a
dipole magnetic field, it is possible that another emission cone, the
"auxiliary cone", is emanating from the other magnetic pole. If it
exists, this auxiliary cone may eventually appear as a new component
in the pulsar profile of A due to geodetic precession. Its appearance,
though, should not affect our predictions for the positioning of the
components of the currently observed cone. In the framework of the
stimulated emission hypothesis, the auxiliary cone will also induce
emission from B. This "secondary" emission should appear at orbital
phases approximately $180^o$ away from those at which enhanced
emission is currently seen. Intriguingly, secondary emission from B
has been reported from these orbital phases$^{11}$, but the enhancement
factor is much smaller than that in the two bright regions. Absorption
or scattering of the radio emission at the shock generated near B by
the interaction between the plasma winds from stars could explain this
effect. A similar shock is believed to cause the eclipses seen in A$^{13}$.

\pagebreak
\psfig{file=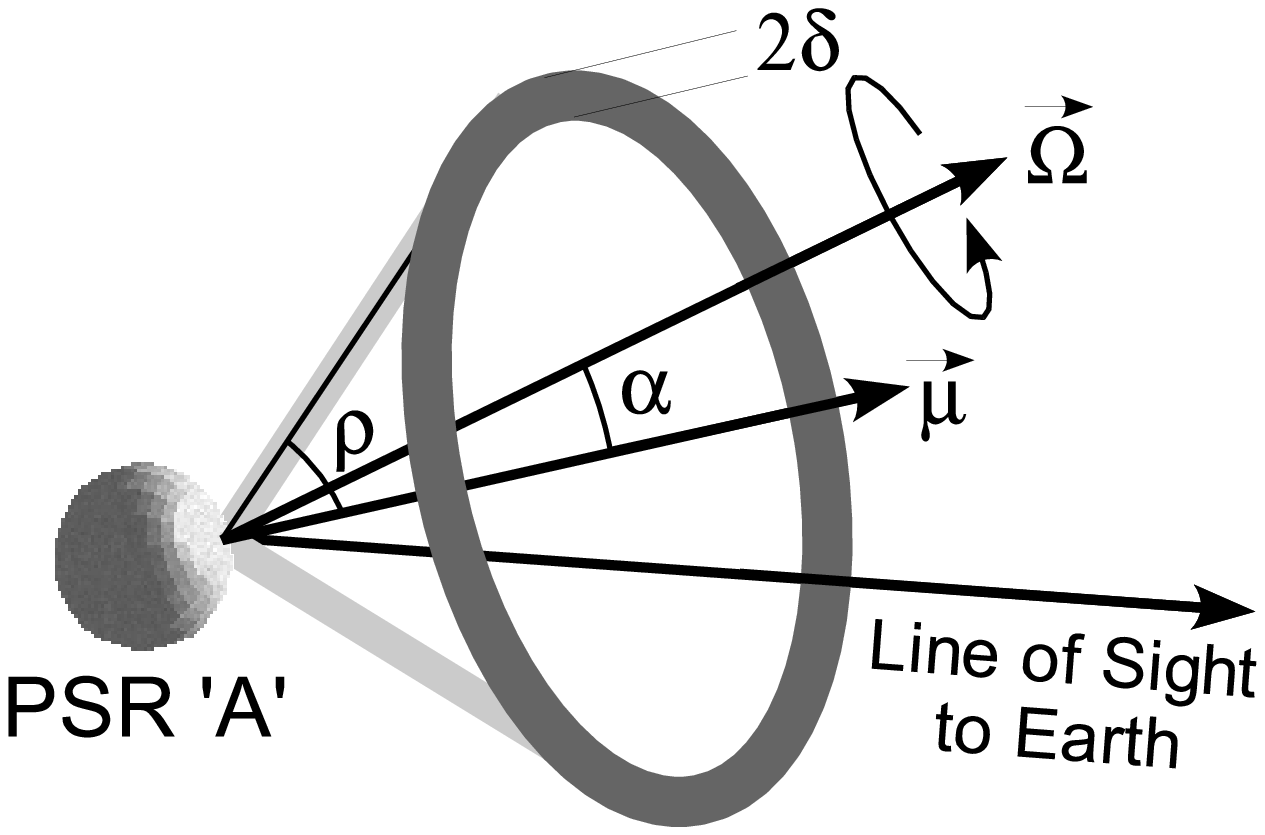,width=5in}
\begin{center} {\large Figure 1} \end{center}

\pagebreak

\psfig{file=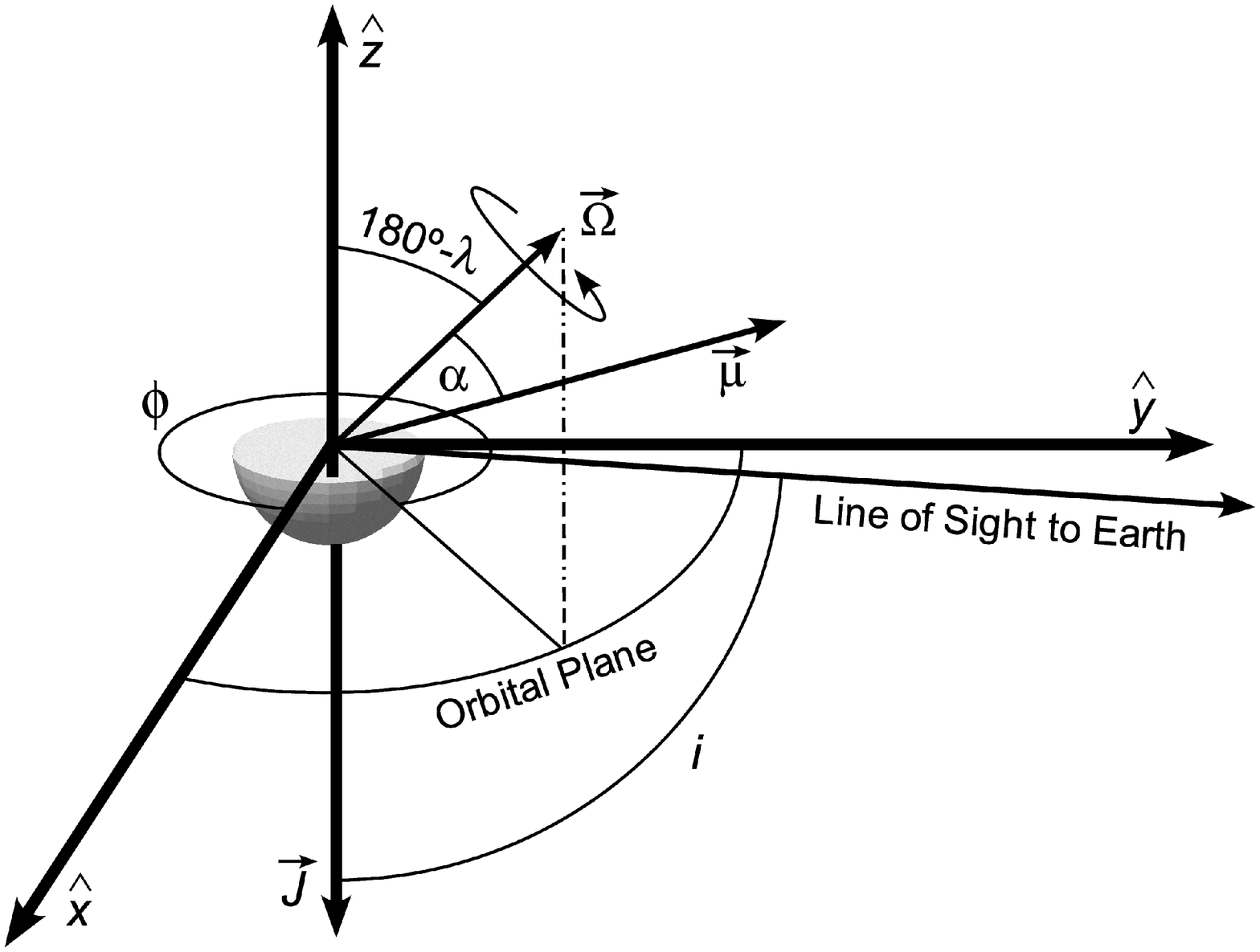,width=5in}
\begin{center} {\large Figure 2} \end{center}

\pagebreak
\psfig{file=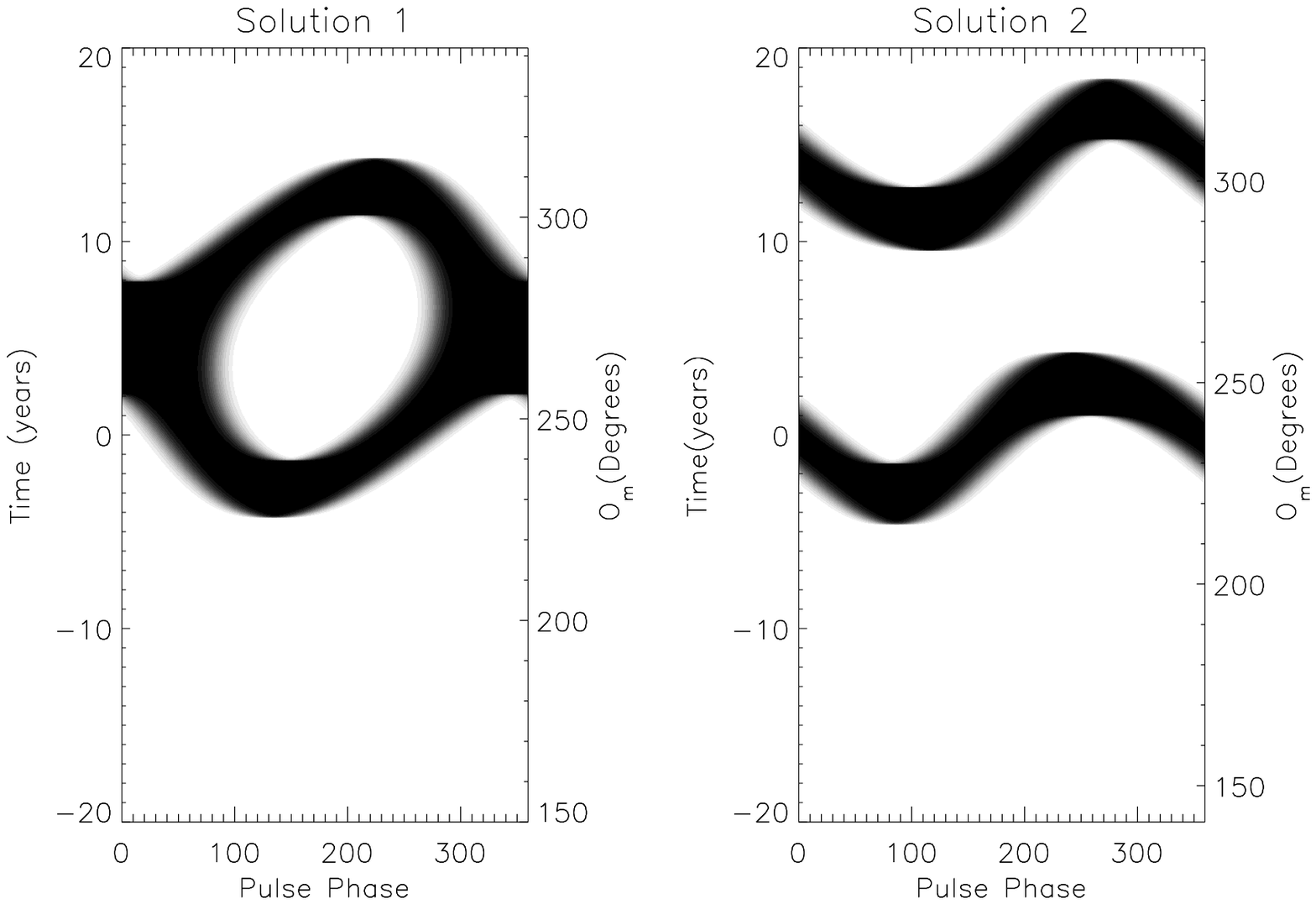,width=5in}
\begin{center} {\large Figure 3} \end{center}

\end{document}